\documentclass[12pt]{article}
\usepackage{amsmath}
\usepackage{amssymb}
\usepackage{amsfonts}
\usepackage{color}

\def\be{\begin{equation}}
\def\ee{\end{equation}}
\def\bea{\begin{eqnarray}}
\def\eea{\end{eqnarray}}
\def\del{\partial}

\def\a{\alpha}
\def\b{\beta}
\def\g{\gamma}

\def\k{\kappa}
\def\p{\psi}
\def\P{\Psi}

\def\d{\delta}
\def\t{\theta}

\def\l{\lambda}
\def\L{\Lambda}
\def\s{\sigma}

\def\hA{\hat{A}}
\def\hD{\hat{D}}
\def\hF{\hat{F}}
\def\hP{\hat{\Psi}}

\def\hL{\hat{\Lambda}}

\def\hd{\hat{\delta}}
\begin{document}
%

\date{}
\title{\vskip-3em\textbf{Seiberg--Witten Maps  to All Orders }}
%
\author{Kayhan \"{U}lker\thanks{E-mail:kulker@gursey.gov.tr.}\,\, and\, Bar\i{}\c{s} Yap\i{}\c{s}kan\thanks{E-mail:yapiskan@gursey.gov.tr}}

\maketitle
\begin{center}
\vskip-3em

\textit{Feza G\"{u}rsey Institute,\\ P.O. Box 6,
TR--34684, \c{C}engelk\"{o}y--Istanbul, Turkey.}

\end{center}

%
\vskip-4em

\begin{abstract}

All order Seiberg--Witten maps of gauge parameter, gauge field and matter fields are given as a closed recursive formula. These maps are obtained by analyzing the order by order solutions of the gauge consistency and equivalence conditions  as well as by directly solving Seiberg-Witten differential equations. The explicit third order non-abelian and fourth order abelian Seiberg-Witten maps of gauge parameter and gauge field are also presented.

\end{abstract}


.

%


\section{Introduction}

The most studied noncommutative (NC) space is the deformation of  D--dimensional Minkowski or Euclidean space $\mathbb{R}^D$ with a real constant antisymmetric parameter $\t$. This NC space can be thought as a (canonical) deformation of the ordinary space-time that is realized with the (Moyal) $*$--product.

Field theories on this NC space can be obtained by replacing the ordinary products with the $*$--products in the expressions.  The idea attracted much attention and studied extensively in the last years mainly due to the fact that NC gauge theory appears as a certain limit of string theory \cite{sw}. This relation to string theory yields the interesting  result that NC gauge theories can be mapped to  commutative ones \cite{sw}. This map is a  gauge equivalence relation between the NC gauge theory and its ordinary counterpart and commonly called as Seiberg-Witten (SW) map.

Such a  relation between NC gauge theory and its ordinary counterpart  can also  be shown to exist by using only the algebraic structure of the aforementioned (canonically deformed) NC space and $*$-product without referring to string theory \cite{mssw,jssw,jmssw}. Moreover, by letting the theory to be an enveloping algebra valued one, one can write the SW map of NC fields for arbitrary non-abelian gauge groups, such as $SU(N)$. Therefore, this progress makes it also possible to construct the Standard Model on the NC space \cite{cjsww}. It is clear that explicit SW--maps of NC fields are needed both to understand the physical predictions and also to check the behavior of the NC theory itself, such as renormalizability.

SW--maps of NC gauge parameter and NC fields can be obtained as solutions of gauge consistency and gauge equivalence conditions respectively \cite{jmssw}. The solutions of these conditions are generally studied perturbatively  by expanding the NC gauge parameter and the NC fields as formal power series in the non-commutativity parameter $\t$.   

A strategy to find these solutions is to determine the dimension and the index structure of the solution first. One then writes the most general expression in terms of fields and their derivatives satisfying these constraints. The coefficients in the solution are fixed by plugging these expressions in the respective equivalence or consistency conditions. This strategy is difficult especially when the higher order solutions in $\t$ are considered. Indeed, the explicit solutions for the SW map of non-abelian gauge theory have been found by various authors only up to second order in $\t$ \cite{jmssw, gh, moller, hakikat, ana, tw}. Due to the freedom in the solutions \cite{ak}, these maps are different from each other up to a homogeneous solution with different coefficients.

Another strategy is to obtain directly solutions of a differential equation  that is generated from the SW--map \cite{sw}. This equation is also commonly called as SW differential equation.  By solving SW differential equation, SW maps of non-abelian gauge parameter and gauge field are given in \cite{wulk} for  a general order-n. SW--map of the matter fields are also obtained in \cite{wulk} for the abelian case. However, the solutions  presented in \cite{wulk} contain explicitly the derivatives with respect to $\t$ and the $*$-product itself. Also they are given as a sum up to order-n. To our best knowledge the relation between the solution of \cite{wulk} and aforementioned second order solutions  were also not studied yet.

In this work, our aim is to write the SW--maps of gauge parameter, gauge field and matter fields of  non-abelian NC YM type theories to all orders. Our starting point is to show that the second order solutions of the gauge field and the gauge parameter given in \cite{moller} can be written in terms of their ordinary counterparts and their first order solutions. This structure of the second order solutions suggest a recursive formula for all orders. We explicitly checked that the third order non-abelian and fourth order abelian solutions stemming from this recursive formula satisfies the SW--map. 

We then show that the same recursive formulas for gauge parameter and gauge field can be obtained from the solutions of the SW differential equation given in \cite{wulk}. Moreover, we  derive the all order SW--maps of matter fields in the fundamental representation of an arbitrary non-abelian group  by solving the respective SW--differential equation. On the other hand, for a field that couples to a gauge field in the adjoint representation of a gauge group, we obtain a similar all order recursive SW--map by simply dimensionally reducing the SW--map of the gauge field.

The organization of the paper is as follows: we review the basics of NC YM theory and SW map in section 2. In section 3, following the original paper \cite{jmssw}, we give the general strategy to solve the NC gauge consistency condition and SW--map for general order--n. We then review the first order  and second order solutions. In particular, we show that the second order solutions given in \cite{moller} can be written in terms of original fields and first order solutions. We then conjecture an all order recursive formula. Combinatoric factors of this recursive formula fixed by checking explicitly the third order non-abelian and fourth order abelian solutions. In section 4, we show that the same recursive formula can also be obtained from the solutions of the SW differential equation given in \cite{wulk}. In section 5, we derive the SW--maps of matter fields to all orders by solving the SW-- differential equation. These maps are also given as closed recursive formulas. We show that the second order solution stemming from this formula coincides with the one of \cite{moller}. The SW--map for the fields in the adjoint representation are also presented in Section 5. In section 6, we end up with conclusions and discussions.


\section{The Seiberg - Witten Map}

NC space that we study in this work is the canonical deformation of the Minkowski space with a real constant antisymmetric parameter $\t^{\mu\nu}$ :
$$[{x} ^\mu \, ,\, {x} ^\nu ]_* \equiv {x} ^\mu \, *\, {x} ^\nu - {x} ^\nu \, *\, {x} ^\mu = i\t^{\mu\nu} .$$
\noindent In this space non--commutativity  is realized with the (Moyal) $*$--product. This $*$--product is associative and due to the fact that the space is deformed with a constant $\t$ it is possible to express it as an exponential :
\bea\label{star}
f(x)*g(x) &\equiv& exp\left(\frac{i}{2}\t^{\mu\nu} \frac{\del}{\del x^{\mu}} \frac{\del}{\del y^{\nu}} \right) f(x)g(y)|_{y\rightarrow x}\\
&=& f(x) \cdot g(x) + \frac{i}{2}\t^{\mu\nu}\del_\mu f(x) \del_\nu g(x) 
-\frac{1}{8} \t^{\mu_1 \nu_1} \t^{\mu_2 \nu_2} \del_{\mu_1}\del_{\mu_2} f(x) \del_{\nu_1}\del_{\nu_2} g(x)+\cdots\nonumber\\
&& + \frac{1}{n!}\left(\frac{i}{2}\right) ^n \t^{\mu_1 \nu_1} \cdots \t^{\mu_n \nu_n} \del_{\mu_1} \cdots \del_{\mu_n} f(x) \del_{\nu_1}\cdots \del_{\nu_n} g(x)   +\cdots  .\nonumber
\eea

NC YM theory is obtained by replacing the ordinary product with the $*$-product (\ref{star}). The NC YM action then reads as\footnote{One $*$-product can be replaced by the ordinary product under the integral since it can be written as a total derivative.} 
\be\label{hI}
\hat{S}=-\frac{1}{4}Tr \int d^4 x \hat{F}^{\mu\nu} * \hat{F}_{\mu\nu} = -\frac{1}{4}Tr \int d^4 x \hat{F}^{\mu\nu}  \hat{F}_{\mu\nu}
\ee
where 
\be\label{hF}
\hat{F}_{\mu\nu} = \del_\mu \hA_{\nu} - \del_\nu \hA_{\mu} - i[\hA_{\mu} , \hA_{\nu}]_* 
\ee
is the NC field strength of the NC gauge field $\hA$. The action (\ref{hI}) is invariant under the NC gauge transformations: 
\be
\hd_{\hL} \hA_{\mu} = \del_{\mu} \hL - i[\hA_{\mu} , \hL]_* \equiv \hD_{\mu}\hL \quad , \quad \hd_{\hL}  \hat{F}_{\mu\nu} = i [\hL , \hat{F}_{\mu\nu}]_* .
\ee
Here, $\hL$ is the NC gauge parameter.

By taking two different limits of string theory, it is possible to define a map (SW--map) between NC fields and their ordinary counterparts as a gauge equivalence relation \cite{sw}: 
\be\label{swe}
\hA_\mu (A;\t) + \hd_{\hL}\hA_\mu (A;\t) = \hA_\mu (A+\d_{\a}A ;\t)  .
\ee
where $A$ and $\a$ are the ordinary (commutative) gauge field and gauge parameter, respectively and $\d_\a $ is the ordinary gauge transformation :
$$
\d_\a A_{\mu} = \del_{\mu} \a - i[A_{\mu} , \a] = D_{\mu}\a  .
$$ 
Eq.(\ref{swe}) can be rewritten as,
\be\label{swe2}
\hd_{\hL}\hA_\mu (A;\t) = \hA_\mu (A+\d_{\a}A ;\t) - \hA_\mu (A;\t) 
= \d_\a \hA_{\mu}(A;\t) .
\ee
The ordinary gauge transformation $\d$ on the r.h.s. of Eq.(\ref{swe2}) acts on the components of $\hA$ when it is expended as a power series in $\t$.

Note that, the NC gauge field $\hA$ and NC gauge parameter $\hL$ have the following functional dependence \cite{sw}: 
\be\label{funcdep}
\hA_\mu = \hA_\mu (A;\t)\quad ,\quad \hat{F}_{\mu\nu} = \hat{F}_{\mu\nu}(A;\t) \quad ,\quad \hL = \hL_\a (\a,A;\t) .
\ee  
Therefore, one has to solve Eq.(\ref{swe}) simultaneously for $\hA_\mu$ and $\hL_\a$ and this is obviously a disadvantage specially when one would like to find the higher order solutions in $\t$.

This difficulty can be disentangled by generalizing the ordinary gauge consistency condition 
$$\d_\a \d_\b - \d_\b \d_\a = \d_{-i[\a ,  \b]}$$ 
to the NC case  \cite{jmssw} :
\be\label{equiv}
i\d_{\a} \hL_\b - i\d_\b \hL_\a - [\hL_\a , \hL_\b]_* = 
i \hL_{-i[\a ,  \b]}  .
\ee
Clearly (\ref{equiv}) is an equation only for the gauge parameter $\hL_\a$ and the solutions can be found order by order.

By constructing NC gauge theory as an enveloping algebra valued one, the map (\ref{swe}) and the  functional dependence (\ref{funcdep}) can be obtained by restricting enveloping algebra valued quantities to depend on their ordinary Lie algebra valued counterparts  \cite{jmssw}.  Moreover, by allowing the theory to be an enveloping algebra valued one, one can construct the NC gauge theory for an arbitrary gauge group like $SU(N)$ \cite{jmssw}.

In the following sections we will not specify the gauge group and our results will be valid for an arbitrary non-abelian gauge group.


\section{Order by Order Solution }

In order to find SW--maps of NC gauge parameter $\hL_\a$ and NC gauge field $\hA_\mu$ one can solve NC gauge consistency condition (\ref{equiv}) and gauge equivalence condition  (\ref{swe2}), respectively, order by order in $\t$ \cite{jmssw}. For this purpose $\hL_\a$ and $\hA_\mu$ can be expanded as formal power series in $\t$,
\bea
\hL_\a &=& \a + \L_\a^{1} + \cdots + \L_\a^{n} + \cdots,\\
\hA_\mu &=& A_\mu + A_{\mu}^{1} + \cdots + A_{\mu}^{n} + \cdots, \nonumber 
\eea
where the zeroth order terms $\a$ and  $A_\mu$  are the ordinary counterparts of $\hL_\a$ and $\hA_\mu$   respectively. The superscript $n$ denotes the order of $\t$ in the expansion.

It is then possible to write gauge consistency equation (\ref{equiv}) as
\be\label{swe-L}
i\d_\a \hL_\b ^n - i\d_\b \hL_\a ^n - \sum_{{p+q+r=n} }  [\L^{p}_\a , \L^{q}_\b]_{*^r}  =
i\hL_{-i[\a ,  \b] } ^n 
\ee
and the gauge equivalence condition (\ref{swe2}) as
\be\label{swe-A}
\d_\a A_{\mu}^{n} = \del_\mu \L^{n}_\a - i \sum_{p+q+r=n}  [A_{\mu}^{p}\, , \,\L^{q}_\a]_{*^r} 
\ee
for the $n$--th order components of $\hL_\a$ and $\hA_\mu$  respectively. Here the sum is over all the values of $p,q$ and $r$ such that ${p+q+r=n}$ and $*^r$ denotes :
$$
f(x)*^{r}g(x) = \frac{1}{r!}\left(\frac{i}{2}\right) ^r \t^{\mu_1\nu_1} \cdots \t^{\mu_r\nu_r} \del_{\mu_1}\cdots\del_{\mu_r}f(x) \del_{\nu_1}\cdots\del_{\nu_r}g(x) .
$$

Following Ref.\cite{moller}, we can rearrange the above  Eq.s (\ref{swe-L}, \ref{swe-A}) for any order $n$ as
\be\label{equiv2}
\Delta {\L}^n \equiv i\d_{\a} \hL_\b ^n - i\d_\b \hL_\a ^n - [{\a} , \hL_\b ^n ] - [\hL_\a ^n , \b] -
i\hL_{-i[\a ,  \b] } ^n =\sum_{\genfrac{}{}{0pt}{}{p+q+r=n,}{p,q \not= n }}  [\L_\a^{p} , \L_\b^{q}]_{*^r}  
\ee
\be\label{equivA}
\Delta_\a A_\mu ^n \equiv \d_\a A_\mu ^n  - i[\a , A_\mu ^n] = \del_\mu \L_\a^n + i \sum_{\genfrac{}{}{0pt}{}{p+q+r=n,}{q \not= n }} [\L_\a^p , A_\mu ^q]_{*^r} 
\ee
so that the l.h.s. contains only the n-th order component of the respective field. 

The strategy to solve the equations is then straightforward: one first constructs the solution $\L_\a^n$ of the Eq. (\ref{equiv2}) at each order and then plugs this solution into (\ref{equivA}) in order to get the solutions $A_\mu^n$ order by order. 

However, to get the n-th order solution, the lower order solutions have to be inserted on the r.h.s. of Eq.s (\ref{equiv2},\ref{equivA}). Therefore, the r.h.s. of these equations explicitly depend on $\t$ and one can extract the homogeneous part from Eq.s (\ref{equiv2}-\ref{equivA}) 
\be\Delta \tilde{\L}^n _\a = 0\quad,\quad \Delta_\a \tilde{A}^n _\mu =0  .
\ee

It is clear that one can add any homogeneous solution $\tilde{\L}^n _\a , \tilde{A}^n _\mu  $ to the inhomogeneous solutions $\L^n _\a , A^n _\mu $ with arbitrary coefficients \cite{jmssw}. This freedom in the solutions were first studied in \cite{ak}.

Note that, the consistency condition (\ref{equiv2}) and gauge equivalence conditions (\ref{equivA}) are given as recursive relations between the lower order solutions and higher order ones. Therefore, it is natural to ask whether the solutions of these equations can also be written in a recursive way or not. 

In the following subsections, we will study these order by order solutions for the NC gauge parameter and NC gauge field  to extract information for the higher order solutions.

\subsection{First order solution : }

We begin with the first order solution given in the original paper \cite{sw} :
\be\label{a1}
\L_\a^{1} = -\frac{1}{4}\t^{\k\l} \{ A_\k ,\del_{\l}\a \} 
\quad ,\quad
A_{\g}^{1} = -\frac{1}{4}\t^{\k\l} \{ A_\k ,\del_\l A_\g + F_{\l\g} \}  .
\ee
One can find the field strength form the definition (\ref{hF}) :
\be\label{f1}
F_{\g\rho}^1 = -\frac{1}{4}\t^{\k\l} \Big( \{ A_\k ,\del_\l F_{\g\rho} + D_\l F_{\g\rho} \} -2 \{ F_{\g\k} , F_{\rho\l}  \} \Big)     .
\ee

We should stress here once more that these solutions are not unique since one can add homogeneous solutions to (\ref{a1}) with arbitrary coefficients. However, as it will be clear in the next subsections, the structure of these solutions will be helpful to obtain the recursive all order solutions.

\subsection{Second order solution : }

The second order solutions that we find useful for our purposes are given in Ref.\cite{moller}. In this set of solutions, the second order gauge parameter $\L^2 _\a$ that satisfies Eq.(\ref{equiv2})
is obtained as \cite{moller}
\bea
\L_\a^{2} &=& {\frac{1}{32}}\t^{\mu\nu}\t^{\k\l}(\{A_\mu,\{\del_\nu
A_\k,\del_\l \a\}\}
+\{A_\mu ,\{A_\k ,\del_\nu \del_\l \a \}\} + \{ \{A_\mu , \del_\nu 
A_\k \} ,\del_\l \a \} \nonumber \\
& -&\{\{ F_{\mu\k} ,A_\nu  \},\del_\l \a \} -2i[\del_\mu  A_\k ,\del_\nu \del_\l \a ] ).\label{ml2}
\eea

The second order gauge field $A^2 _\mu$ can then be found\footnote{Indeed, the solution for gauge field $A_{\mu}^{2}$ given in \cite{moller} has typographic errors. The correct solution with correct signs is given in \cite{hakikat}.} by inserting the above solution (\ref{ml2}) in the Eq.(\ref{equivA})  :
\bea\label{ma2}
A_{\g}^{2}&=& \frac{1}{32}\t^{\mu\nu} \t^{\k \l} \Big(\{\{A_\k,\del_\l
A_\mu\} ,\del_\nu A_\g \}-\{\{F_{\k\mu},
A_\l \},\del_\nu A_\g \}-2i[\del_\k
A_\mu ,\del_\l \del_\nu A_\g ] \\
&& -\{A_\mu, \{\del_\nu F_{\k \g},
A_\l \} \} - \{A_\mu, \{ F_{\k \g}, \del_\nu
A^0_\l \}\}  +
\{A_\mu , \{\del_\nu A_\k , \del_\l A_\g \} \} \nonumber\\
&& + \{A_\mu, \{ A_\k, \del_\nu \del_\l
A_\g \} \} - \{\{A_\k,\del_\l F_{\mu\g} \}, A_\nu\}
+\{\{D_\k  F_{\mu\g} , A_\l \}, A_\nu \} \nonumber \\
&& +2\{\{F_{\mu\k} ,F_{\g\l } \},
A_\nu \} + 2i [\del_\k F_{\mu\g} ,\del_\l
A_\nu ]-\{F_{\mu\g} ,\{ A_\k,
\del_\l A_\nu \}\}+\{F_{\mu\g },\{F_{\k
  \nu} , A_\l \} \}\Big)   .  \nonumber
\eea 

One can explicitly check that (\ref{ma2}) together with (\ref{ml2}) satisfies gauge equivalence relation (\ref{equivA}).

The reason to begin with the solutions (\ref{ml2},\ref{ma2}) becomes apparent now:   (\ref{ml2},\ref{ma2}) can be written in terms of lower order solutions after rearranging the indices and regrouping the relevant terms together :
\be
\L_\a^{2} = -\frac{1}{8}\t^{\k\l}\left( \{A_{\k}^{1}, \del_{\l}\a \} 
+ \{A_{\k}, \del_{\l}\L_\a^{1}\} \right)
-\frac{i}{16}\t^{\k\l}\t^{\mu\nu}[\del_\mu A_\k , \del_\nu \del_{\l}\a ] \label{l2}
\ee
\bea \label{a2}
A_{\g}^{2} &=& -\frac{1}{8}\t^{\k\l}\left( \{A_{\k}^{1}, \del_\l A_\g + F_{\l\g} \} 
+ \{A_{\k}, \del_\l A_\g ^1 + F_{\l\g} ^{1}\} \right) \\ 
&& -\frac{i}{16}\t^{\k\l}\t^{\mu\nu}[\del_\mu A_\k , \del_\nu (\del_\l A_\g + F_{\l\g} ) ] . \nonumber
\eea

The field strength at the second order given in \cite{moller} can also be written in terms of first order solutions :
\bea
F_{\g\rho} ^2 &=& -\frac{1}{8}\t^{\k\l} \Big( \{ A_\k ,\del_\l F_{\g\rho} ^1 + (D_\l F_{\g\rho} )^1  \} + \{ A_\k ^1 ,\del_\l F_{\g\rho}  + D_\l F_{\g\rho} \} \\
&& \qquad\qquad  -2 \{ F_{\g\k} , F_{\rho\l} ^1 \} -2 \{ F_{\g\k} ^1 , F_{\rho\l}  \} \Big) \nonumber\\
&& -\frac{i}{16}\t^{\k\l}\t^{\mu\nu} \Big( [ \del_\mu A_\k ,\del_\nu \big( \del_\l F_{\g\rho} + D_\l F_{\g\rho} \big) ] -2 [ \del_\mu F_{\g\k} , \del_\nu F_{\rho\l} ]   \Big) . \nonumber
\eea
Here,  $(D_\l F_{\g\rho} )^1 $ denotes
$$
(D_\l F_{\g\rho} )^1 =  D_\l F_{\g\rho} ^1 -i [A^1 _\l , F_{\g\rho}] +\frac{1}{2}\t^{\mu\nu} \{\del_\mu A _\l , \del_\nu F_{\g\rho} \} .
$$

Obviously, when the solutions are written in the above form, it is much easier to show that they satisfy the gauge equivalence condition (\ref{equivA}).

On the other hand, these solutions are not the most general solutions. For the second order in $\t$, several different solutions are obtained by various authors in \cite{gh,jmssw,moller,hakikat} and recently in \cite{ana,tw}. All these solutions are related to each other up to addition of homogeneous solutions with different coefficients. Therefore, for the most general solution one should also include the homogeneous solutions. However, inclusion of the homogeneous solutions with arbitrary coefficients makes it difficult to see how to write these solutions in a recursive manner and hence to all orders.

\subsection{ n--th order solutions : }

When  the structure of the first order solutions (\ref{a1}) are compared with the second order ones  (\ref{l2},\ref{a2}), one sees immediately that the terms linear in $\t$ on the r.h.s. of (\ref{l2},\ref{a2}) come from an expansion whose first order term can be written as (\ref{a1}). Quadratic terms in $\t$ are clearly related with the expansion of the $*$-product itself. Therefore, by analyzing first two order solutions one can conjecture the general structure:  
\be\label{nl}
\L^{n+1} _\a = -\frac{1}{4(n+1)}\t^{\k\l}\sum_{{p+q+r=n}} \{A_\k ^p , \del_\l \L^q _\a \}_{*^r}
\ee
\be\label{na}   
A_{\g}^{n+1} = -\frac{1}{4(n+1)}\t^{\k\l}\sum_{{p+q+r=n}} \{ A_{\k}^{p}, \del_\l A^q_\g + F^{q}_{\l\g} \}_{*^{r}}   .
\ee 

By following similar steps we can also write the n-th order term of the field strength
\be\label{nf}
F_{\g\rho} ^n = -\frac{1}{4(n+1)}\t^{\k\l}\sum_{{p+q+r=n}} \big( \{A_\k ^p ,\del_\l F_{\g\rho} ^q + (D_\l F_{\g\rho} )^q \} -2 \{ F_{\g\k} ^p , F_{\rho\l} ^q \}_{*^r} \big)
\ee
where 
$$
(D_\l F_{\g\rho} )^n  = \del_\l F_{\g\rho} ^n -i \sum_{{p+q+r=n}} [A^p _\l , F_{\g\rho} ^q]_{*^r}.
$$

In most cases, the first two order terms in an expansion  is not sufficient to write down a general formula. However, most of the time the structure of the third order term suggests strongly the general n-th order term with the correct combinatoric factors. Therefore,  to show the validity of our conjecture  and to check the overall constant given in (\ref{nl},\ref{na}), we write down the conjectured third order non-abelian solutions explicitly: \bea\label{l3}
\L_\a^{3} &=& -\frac{1}{12}\t^{\k\l}\left( \{A_{\k}^{2}, \del_{\l}\a \} 
+ \{ A_{\k}^{1}, \del_{\l}\L_\a^{1}\} + \{A_{\k}, \del_{\l}\L_\a^{2}\} \right) \\
&&
-\frac{i}{24}\t^{\k\l}\t^{\mu\nu}\left( [\del_\mu A_{\k}^{1} , \del_\nu \del_{\l} \a ] + [\del_\mu A_\k , 
\del_\nu \del_{\l}\L_\a^{1} ] \right) \nonumber\\
&&+ \frac{1}{96} \t^{\k\l} \t^{\mu\nu} \t^{\rho\s}  \{ \del_\mu \del_\rho  A_{\k} , 
\del_\l \del_\nu \del_\s \a \}   .   \nonumber 
\eea

Indeed, one can show that third order solution (\ref{l3}) for the gauge parameter $\L^3 _\a$ satisfies the gauge consistency condition (\ref{equiv2}). We also checked explicitly that the conjectured third order gauge field $A^3 _\mu$ 
\bea\label{a3}
A_{\g}^{3} &=& -\frac{1}{12}\t^{\k\l}\big( \{A_{\k}^{2}, \del_\l A_\g + F_{\l\g} \} 
+ \{ A_{\k}^{1}, \del_\l A_\g ^1 + F_{\l\g}^{1}\} + \{A_{\k}, \del_\l A_\g ^2 + F_{\l\g}^{2}\} \big) \\
&& -\frac{i}{24}\t^{\k\l}\t^{\mu\nu}\big( [\del_\mu A_{\k}^{1} , \del_\nu (\del_\l A_\g + F_{\l\g} ) ] 
+ [\del_\mu A_\k , \del_\nu (\del_\l A_\g ^1 + F_{\l\g}^{1} ) ] \big) \nonumber\\
&&+ \frac{1}{96} \t^{\k\l} \t^{\mu\nu} \t^{\rho\s}  \{ \del_\mu \del_\rho  A_{\k} , 
\del_\nu \del_\s (\del_\l A_\g + F_{\l\g} ) \} \nonumber 
\eea
together with $\L^3 _\a$ given in (\ref{l3}) satisfies the SW--map (\ref{equivA}). The overall constant $-{1}/{4(n+1)}$  is fixed uniquely with these third order solutions.

The fourth order SW map can be written from (\ref{nl},\ref{na}) as  
\bea\label{l4}
\L^{4} _\a  &=& -\frac{1}{16}\t^{\k\l}\big( \{ A_{\k}^{3}, \del_{\l}\a \} 
+ \{ A_{\k}^{2}, \del_{\l}\L_\a^{1} \} + \{ A_{\k}^{1}, \del_{\l}\L_\a^{2} \}  + \{ A_{\k}, \del_{\l}\L_\a^{3} \} \big) \\
&& -\frac{i}{32}\t^{\k\l}\t^{\mu\nu}\big( [ \del_\mu A_{\k}^{2},  \del_\nu \del_{\l} \a ] + [ \del_\mu A_{\k} ^1,  \del_\nu \del_{\l}\L_\a^{1} ] + [ \del_\mu A_{\k} ,  \del_\nu \del_{\l}\L_\a^{1} ]  \big)\nonumber\\ 
&& + \frac{1}{128} \t^{\k\l} \t^{\mu\nu} \t^{\rho\s} \big( \{ \del_\mu \del_\rho  A_{\k} ^1, \del_\s \del_\nu \del_\l \a \} + \{ \del_\mu \del_\rho  A_{\k}, \del_\s \del_\nu \del_\l \L_\a^1 \} \big) \nonumber\\ 
&& + \frac{i}{768} \t^{\k\l} \t^{\mu\nu} \t^{\rho\s} \t^{\zeta\eta} [ \del_\zeta \del_\rho \del_\mu A_\k, \del_\eta \del_\s \del_\nu \del_\l \a ]\nonumber
\eea
and 
\bea\label{a4}
A_{\g}^{4} &=& -\frac{1}{16}\t^{\k\l}\Big( \{ A_{\k}^{3}, \del_\l A_\g  + F_{\l\g} \} 
+ \{ A_{\k}^{2}, \del_\l A_\g ^1 + F_{\l\g}^{1} \} \\
&& \qquad \qquad + \{ A_{\k}^{1}, \del_\l A_\g ^2 + F_{\l\g}^{2} \} 
+ \{ A_{\k}, \del_\l A_\g ^3 + F_{\l\g}^{3} \} \Big)\nonumber \\
&& -\frac{i}{32}\t^{\k\l}\t^{\mu\nu}\Big( [ \del_\mu A_{\k}^{2},  \del_\nu ( \del_\l A_\g  + F_{\l\g}  ) ]
+ [ \del_\mu A_{\k}^{1},  \del_\nu ( \del_\l A_\g ^1 + F_{\l\g} ^1  ) ] \nonumber \\
&& \qquad \qquad \qquad+ [ \del_\mu A_{\k}, \del_\nu ( \del_\l A_\g ^{2}  + F_{\l\g} ^{2} ) ]  \Big) \nonumber\\
&& + \frac{1}{128} \t^{\k\l} \t^{\mu\nu} \t^{\rho\s} \Big(  \{ \del_\mu \del_\rho  A_{\k} ^1, 
\del_\nu \del_\s ( \del_\l A_\g   + F_{\l\g}  ) \}   + \{ \del_\mu \del_\rho  A_{\k},
\del_\nu \del_\s ( \del_\l A_\g ^1  + F_{\l\g} ^1  ) \} \Big) \nonumber\\ 
&& + \frac{i}{768} \t^{\k\l} \t^{\mu\nu} \t^{\rho\s} \t^{\zeta\eta} [ \del_\zeta \del_\rho \del_\mu A_\k, 
\del_\eta \del_\s \del_\nu ( \del_\l A_\g + F_{\l\g} ) ]   . \nonumber
\eea

One can plug (\ref{l4},\ref{a4}) in (\ref{equivA}) to check the validity of our conjecture. However, the calculations are cumbersome even though (\ref{l4},\ref{a4}) are written in a recursive way.

Nevertheless, it is still manageable to calculate for the abelian case since all the commutator terms vanish. Therefore, to check further the above result, we also carried our calculation to the fourth order abelian case that can be obtained easily from (\ref{l4},\ref{a4}) by setting all the commutators to zero. We verified that these abelian solutions satisfy the SW--map (\ref{swe2}). 

We may conclude already at this level that the aforementioned results strongly suggests that the above given conjecture   (\ref{nl}, \ref{na}) is valid for all order solutions.


\section{Solution of Seiberg-Witten Differential Equation : }

In the previous section, by analyzing the order by order solutions we could write down a recursive all order solution. However, it is possible to find the same solution from a differential equation introduced in the original paper \cite{sw}. This equation is also often called SW-differential equation and
obtained by varying the deformation parameter infinitesimally $\t \rightarrow \t +\d\t $ \cite{sw}:
\be
\d\t^{\mu\nu}\frac{\del\hA_\g}{\del\t^{\mu\nu}} = -\frac{1}{4}\d\t^{\k\l} \{\hA_\k ,\del_\l \hA_\g + \hF_{\l\g}\}_*
\ee 
which can be written as 
\be\label{difeq}
\frac{\del\hA_\g}{\del\t^{\k\l}} =  -\frac{1}{8}\{\hA_\k ,\del_\l \hA_\g + \hF_{\l\g}\}_*  + \frac{1}{8} \{\hA_\l ,\del_\k \hA_\g + \hF_{\k\g}\}_* .
\ee 

The solution of the SW differential equation (\ref{difeq}) is obtained in Ref.\cite{wulk} as 
\be\label{hl}
\hL^{(n+1)} _\a = \a -\frac{1}{4} \sum_{k=1}^{n+1}\frac{1}{k!}\t^{\mu_1\nu_1}\t^{\mu_2\nu_2} \cdots \t^{\mu_k\nu_k} \left( \frac{\del^{k-1}}{\del\t^{\mu_2\nu_2} \cdots \del\t^{\mu_k\nu_k} } \{\hA^{(k)}_{\mu_1},\del_{\nu_1}\hL^{(k)} _\a  \}_* \right)_{\t =0}
\ee
\bea\label{ha}
&&\hA^{(n+1)}_\g =A_\g \\
&& -\frac{1}{4} \sum_{k=1}^{n+1}\frac{1}{k!}\t^{\mu_1\nu_1}\t^{\mu_2\nu_2} \cdots \t^{\mu_k\nu_k} \left( \frac{\del^{k-1}}{\del\t^{\mu_2\nu_2} \cdots \del\t^{\mu_k\nu_k} } \{\hA^{(k)}_{\mu_1}, \del_{\nu_1} \hA^{(k)} _\g + \hat{F}^{(k)}_{\nu_1 \g}  \}_* \right)_{\t =0} .\nonumber
\eea
after expanding the NC gauge parameter and NC gauge field into a Taylor series\footnote{We refer to the original paper \cite{wulk} for the details.}. Here, $\hL_\a^{(n)}$ and $A_\mu^{(n)}$ denotes the sum up to order $n$ :
\bea
\hL_\a ^{(n)}&=& \a + \L_\a^{1} + \cdots + \L_\a^{n} ,\nonumber\\
\hA_\mu ^{(n)} &=& A_\mu + A_{\mu}^{1} + \cdots + A_{\mu}^{n}  . \nonumber 
\eea

Note that, contrary to the solutions presented in the previous section, these solutions explicitly contain the derivatives w.r.t. $\t$ and the $*$--product itself and given as a sum of all (n+1) orders. However, it is possible to extract the recursive solutions (\ref{nl},\ref{na}) from the sum (\ref{hl},\ref{ha}).

For this purpose let us write the $n+1$--st component of  (\ref{hl}) :
\be
\L^{n+1} _\a =  -\frac{1}{4(n+1)!}\t^{\mu\nu}\t^{\mu_1\nu_1} \cdots \t^{\mu_n\nu_n} \left( \frac{\del^{n}}{\del\t^{\mu_1\nu_1} \cdots \del\t^{\mu_n\nu_n} } \{\hA^{(n)}_{\mu_1},\del_{\nu_1} \hat{\L}^{(n)} _\a  \}_*
 \right)_{\t =0}  .
\ee

Since, $\t$ is set to zero after taking the derivatives, the expression in the paranthesis can be written as a sum up to n-th order: 
\be\label{wln}
\L^{n+1} _\a =  -\frac{1}{4(n+1)!}\t^{\mu\nu}\t^{\mu_1\nu_1} \cdots \t^{\mu_n\nu_n} \left( \frac{\del^{n}}{\del\t^{\mu_1\nu_1} \cdots \del\t^{\mu_n\nu_n} } \sum_{p+q+r=n} \{ A_{\mu}^{p}, \del_\nu \L^{q} _\a  \}_{*^{r}} \right)  .
\ee

It is then an easy exercise to show that the equation (\ref{wln}) reduces to the recursive formula (\ref{nl}) given in the previous section :
$$
\L^{n+1} _\a =  -\frac{1}{4(n+1)}\t^{\mu\nu} \sum_{p+q+r=n} \{ A_{\mu }^{p}, \del_{\nu }\L^{q} _\a \}_{*^{r}} .
$$

With the same algebraic manipulation one can also derive the same recursive formula for the gauge field (\ref{na}) 
\bea
A^{n+1}_\g &=&  -\frac{1}{4(n+1)!}\t^{\mu\nu}\t^{\mu_1\nu_1} \cdots \t^{\mu_n\nu_n} \left( \frac{\del^{n}}{\del\t^{\mu_1\nu_1} \cdots \del\t^{\mu_n\nu_n} } \{\hA^{(n)}_{\mu_1},\del_{\nu_1} \hA^{(n)} _\g + \hat{F}^{(n)}_{\nu_1 \g}  \}_*
 \right)_{\t =0}\\
&=&  -\frac{1}{4(n+1)}\t^{\mu\nu} \sum_{p+q+r=n} \{ A_{\mu }^{p},  \del_\nu A^q _\g + F^{q}_{\nu\g} \}_{*^{r}} . \nonumber
\eea


\section{Seiberg--Witten Map for Matter Fields }

SW--map of a NC field  $\hP$ that couples to a gauge field $\hA$ in a gauge invariant theory can be derived from a similar gauge equivalence relation \cite{jmssw} :
\be\label{swe3}
\hd_{\hL}\hP(\p , A;\t) =  \d_\a \hP(\p ,A;\t) .
\ee
such that the NC field $\hP$  is also a functional of its ordinary counterpart $\p$ and the gauge field $A$. After expanding the NC field $\hP$ as formal power series in $\t$
\be
\hP = \p + \hP^{1} + \cdots + \hP^{n} + \cdots
\ee
the solution of (\ref{swe3}) can be found order by order by using a similar method revised in Section 2.

The equivalence relation (\ref{swe3}) is valid both for  bosonic and  fermionic fields although obviously their corresponding kinetic and interaction terms in an action are different. Therefore, in the following discussions we will not distinguish the nature of the field $\hP$, since for both cases the structure of the equivalence relation (\ref{swe3}) and hence the structure of the solutions are the same.

On the other hand, the equivalence relation (\ref{swe3}) is also valid for $\p$ either in the fundamental representation or in the adjoint  representation of an arbitrary non-abelian gauge group. We will derive all order recursive solutions of  (\ref{swe3}) for both of the cases by using different approaches.

\subsection{Fundamental Representation}

When the ordinary field $\p$ is in the fundamental representation the gauge transformation reads as 
\be\label{gfund}
\d_\a \p = i \a \p .
\ee
The NC generalization of the gauge transformation of a NC field $\hP$ is defined by replacing the ordinary product with $*$--product :
\be
\hd_{\hL} \hP = i \hL_\a * \hP .
\ee   

Following the general strategy presented in section 2, we can write the gauge equivalence relation  (\ref{swe3}) as  
\be\label{equivP}  
\Delta_\a \P ^n \equiv \d_\a \P ^n  - i \a  \P ^n  = i \sum_{\genfrac{}{}{0pt}{}{p+q+r=n,}{q \not= n }} 
  \L_\a^p {*^r} \P ^q   ,
\ee
for all orders. As discussed before, the solution of this equation is not unique. One is free  to add any homogeneous solution $\tilde{\P}^n$ of the equation 
$$\Delta_\a \tilde{\P}^n=0$$
to the solutions ${\P}^n$ of (\ref{equivP}) \cite{jmssw}.

A solution of Eq.(\ref{equivP}) for the first order is given in Ref.\cite{jmssw} :
\be\label{p1}
\P^1  = -\frac{1}{4}\t^{\k\l} A_\k  (\del_\l + D_\l)\p 
\ee
where $D_\mu$ is the covariant derivative :
$$
D_\mu \p = \del_\mu \p -i A_\mu \p \,.
$$

The second order solutions of (\ref{equivP}) are also studied by several authors \cite{jmssw,moller,hakikat,ana,tw}. A natural guess to write the solution (\ref{equivP})  in a recursive way in terms of the ordinary fields $A_\mu$ and $\p$ and their first order solutions $A_\mu^1$ and $\P^1$ is to use the solution given in \cite{moller}. However, for matter fields it is not easy to see this structure at first sight. 

Nevertheless one can derive the SW differential equation from the first order solution and find its solutions. For this purpose by following Ref.\cite{sw}, we derive  the differential equation from (\ref{p1}) by varying the deformation parameter infinitesimally $\t \rightarrow \t + \d\t$ : 
\be\label{difeqfp}
\d\t^{\mu\nu}\frac{\del\hP}{\del\t^{\mu\nu}}  = -\frac{1}{4}\d\t^{\k\l}  \hA_\k  * (\del_\l \hP + \hD_\l \hP )
\ee
which can also be written as     
\be\label{difeqp}
\frac{\del\hP}{\del\t^{\k\l}}  = -\frac{1}{8}  \hA_\k * (\del_\l \hP + \hD_\l \hP)  +\frac{1}{8} \hA_\l  * (\del_\k \hP + \hD_\k \hP) 
\ee
The NC covariant derivative here is defined as :
$$
\hat{D}_\mu \hP = \del_\mu \hP -i  \hA_\mu * \hP \,.
$$

By using a similar method presented in \cite{wulk},  we expand $\hP$ in Taylor series :
\bea
\hP^{(n+1)} &=& \p + \P^{1} + \P^{2} + \cdots + \P^{n+1} 
 \\  
&=& \p +\sum_{k=1}^{n+1}\frac{1}{k!}\t^{\mu_1\nu_1}\t^{\mu_2\nu_2} \cdots \t^{\mu_k\nu_k} \left( \frac{\del^{k}}{\del\t^{\mu_1\nu_1} \cdots \del\t^{\mu_k\nu_k} } \left(\hP^{(n+1)}\right)\right)_{\t =0}. \nonumber
\eea

The solution of the differential equation (\ref{difeqp}) can then be found as
\be\label{HP}
\hP^{(n+1)} =  \p -\frac{1}{4} \sum_{k=1}^{n+1}\frac{1}{k!}\t^{\mu_1 \nu_1}\t^{\mu_2 \nu_2} \cdots \t^{\mu_k \nu_k} \big( \frac{\del^{k-1}}{\del\t^{\mu_2 \nu_2} \cdots \del\t^{\mu_k \nu_k} }  \hA_{\mu_1} ^{(k)} * (\del_{\nu_1} \hP ^{(k)} + (\hD_{\nu_1} \hP)^{(k)}) \big)_{\t =0}
\ee
where
$$
(\hD_\mu \hP)^{(n)} = \del_\mu \hP^{(n)} -i \hA _\mu ^{(n)} * \hP^{(n)} .
$$

For the abelian case this solution  differs from the one given in \cite{wulk} by a homogeneous solution.

Following the similar steps presented in Section 5, to find the all order recursive solution of the gauge equivalence relation (\ref{equivP})  we write the $n+1$--st component of (\ref{HP}) explicitly :
\bea
\P^{n+1} &=&  -\frac{1}{4(n+1)!}\t^{\mu\nu}\t^{\mu_1\nu_1} \cdots \t^{\mu_n\nu_n} \left( \frac{\del^{n}}{\del\t^{\mu_1\nu_1} \cdots \del\t^{\mu_n\nu_n} }  \hA_\mu ^{(n)} * (\del_\nu \hP ^{(n)} + (\hD_\nu \hP)^{(n)})
 \right)_{\t =0}\\
&=&  -\frac{1}{4(n+1)!}\t^{\mu\nu}\t^{\mu_1\nu_1} \cdots \t^{\mu_n\nu_n} \left( \frac{\del^{n}}{\del\t^{\mu_1\nu_1} \cdots \del\t^{\mu_n\nu_n} } \sum_{p+q+r=n} A_\mu ^{p} {*^{r}} (\del_\nu \P ^{(q)} + (D_\nu \P)^{q}) \right) \nonumber
\eea
where
$$
(D_\mu \P)^{n} = \del \P^n -i \sum_{p+q+r=n} A_\mu ^p *^r \P^q .
$$

After taking derivatives w.r.t. $\t$'s we obtain the all order recursive solution of the gauge equivalence relation  (\ref{equivP}) :
\be\label{np}
\p^{n+1} = -\frac{1}{4(n+1)}\t^{\k\l} \sum_{p+q+r=n} A_\k ^{p} {*^{r}} (\del_\l \P ^{(q)} + (D_\l \P)^{q})   .
\ee 

It is easy to see that the first order solution (\ref{p1}) is obtained by setting $n=0$. By setting $n=1$ the second order solution reads  :
\bea\label{p2}
\p^2 &=& -\frac{1}{8}\t^{\k\l} \big( 2A_\k ^1 \del_\l \p - i A_\k ^1 A_\l \p + 
2A_\k \del_\l \p ^1 -i A_\k A_\l ^1 \p - iA_\k A_\l \p ^1 \\
&&+ i \t^{\mu\nu}\del_\mu A_\k  \del_\nu \del_\l \p 
 + \frac{1}{2} \t^{\mu\nu}\del_\mu A_\k  \del_\nu A_\l  \p 
+ \frac{1}{2} \t^{\mu\nu}\del_\mu A_\k   A_\l \del_\nu  \p             
+ \frac{1}{2} \t^{\mu\nu} A_\k \del_\mu   A_\l \del_\nu  \p \big) .   \nonumber
\eea
One can check that this solution satisfies Eq.(\ref{equivP}).

On the other hand, to compare the second order solution (\ref{p2}) with the other solutions in the literature, we insert the first order solutions (\ref{a1},\ref{p1}) in (\ref{p2}) :
\bea\label{mp2}
\p^2
= \frac{1}{32} \t^{\mu\nu} \t^{\k\l} \big( && -4i \del_\mu A_\k \del_\l \del_\nu \p
+ 4 A_\mu A_\k \del_\l \del_\nu \p - 4 \del_\mu A_\k A_\nu \del_\l \p - 4 A_\mu \del_\k 
A_\nu \del_\l \p \\ \nonumber 
&& + 8 A_\mu \del_\nu A_\k \del_\l \p - 2\del_\mu A_\k 
\del_\nu A_\l \p 
+ 4A_\mu A_\k A_\nu \p - 3A_\mu A_\nu A_\k A_\l \p \\ \nonumber && - 2A_\mu A_\k A_\l A_\nu \p
+ 4i A_\mu A_\k A_\nu \del_\l \p - 4i A_\mu A_\k A_\l \del_\nu \p - 4iA_\mu A_\nu A_\k \del_\l \p
\\ \nonumber && +2i \del_\mu A_\k A_\nu A_\l \p - 2i A_\mu A_\k \del_\l A_\nu \p - i\del_\mu 
A_\k A_\l A_\nu \p - 5i A_\mu \del_\nu A_\k A_\l \p \\ \nonumber 
&& + 3i A_\mu \del_\k A_\nu A_\l \p - i A_\mu A_\k \del_\nu A_\l \p \big)      .
\eea

After regrouping the relevant terms in (\ref{mp2}), this solution gives exactly the one presented in Ref.\cite{moller}.

\subsection{Adjoint representation}     

When the ordinary field $\p$ is in the adjoint representation the gauge transformation reads as 
\be\label{gadj}
\d_\a \p = i [ \a , \p ]
\ee
Non--commutative generalization of the gauge transformation  (\ref{gadj}) can then be written as
\be\label{hgadj}
\hat{\d}_{\hL} \hP = i [ \hL_\a , \hP ]_* \, .
\ee

Following the general strategy, the gauge equivalence relation (\ref{swe3}) can be written as
\be\label{equivaP}  
\Delta_\a \P ^n := \d_\a \P ^n  - i [ \a , \P ]  = i \sum_{\genfrac{}{}{0pt}{}{p+q+r=n,}{q \not= n }} 
  [\L_\a^p , \P ^q ]_{{*^r}}  
\ee
for all orders.

The solutions of Eq.(\ref{equivaP}) can be found either by directly solving this equation order by order or by solving the respective differential equation as discussed in the previous sections. However, when the general solution of the equivalence relation for the gauge field (\ref{equivA}) is known, possibly the easiest way to obtain the solution is to use dimensional reduction. Indeed, the first order solution is given in \cite{su} via dimensional reduction\footnote{See also Ref.\cite{by}} :
\be
\p^{1} = -\frac{1}{4} \t^{\k\l} \{ A_\k  ,  \del_\l \p  + D_\l \p \}    .
\ee

Therefore, by setting the components of the deformation parameter $\t$ on the compactified dimensions zero, the trivial dimensional reduction\footnote{We refer to Ref.\cite{su} for  details.} (i.e. from six to four dimensions) of (\ref{na}) leads to the general n--th order solution for a complex scalar field :
\be\label{nap}
\p^{n+1} = -\frac{1}{4(n+1)}\t^{\k\l} \sum_{p+q+r=n} \{ A_\k ^{p}  ,  (\del_\l \P ^{(q)}  + (D_\l \P)^{q}) \}_{*^{r}}   .
\ee
Since the structure of the solutions are the same for both the scalar and fermionic fields, this solution (\ref{nap}) can also be used for the fermionic fields.

Note that, after introducing the anticommutators/commutators properly, the form of the solution (\ref{nap}) is similar with the one (\ref{np}) given in the previous section except that the covariant derivative is now given as
$$
D_\mu \p =\del_\mu \p -i [A_\mu , \p] 
$$
and hence
$$
(D_\mu \P )^n =\del_\mu \P^n -i \sum_{p+q+r=n} [A_\mu ^p , \P^q ]_{*^r} \,  .
$$

The same result can also be obtained by solving the differential equation :
\be
\frac{\del\hP}{\del\t^{\k\l}}  = -\frac{1}{8}  \{ \hA_\k  ,(\del_\l \hP + \hD_\l \hP)\}_*  +\frac{1}{8} \{ \hA_\l  ,  (\del_\k \hP + \hD_\k \hP)\}_*  . 
\ee
Therefore, the result  obtained above via dimensional reduction (\ref{nap}) can also be thought  as an independent check of the aforementioned results.


\section{Conclusion}

In this work we presented all order SW--maps (\ref{nl},\ref{na},\ref{np}) of gauge parameter, gauge field and matter fields respectively. These maps are given as closed recursive expressions. It is also  possible to write these maps explicitly in terms of ordinary fields and gauge parameter to any desired order with a little effort. Other solutions can be obtained by adding homogeneous solutions of (\ref{equiv2},\ref{equivA},\ref{equivP}) at any order.  

These maps are obtained by using different methods. One is to  examine the solutions of gauge consistency conditions (\ref{equiv2}) and gauge equivalence conditions (\ref{equivA},\ref{equivP}) order by order. We showed that the solutions given in \cite{moller} up to second order in $\t$ admits a recursive formulation in terms of first order solutions and the original fields. This structure leads to the aforementioned all order SW--maps.  The other method is to solve directly the respective SW differential equations. We showed that these solutions also lead to the same all order SW--maps.

It is worth mentioning that the SW differential equation (\ref{difeq}) can also be obtained in an elegant way by studying the BRST cohomology of the NC YM theory \cite{bbg1,bbg2}. In this setting, it is shown that the most general solution including the homogeneous solutions  of Eq.(\ref{difeq}) can be obtained recursively at any order \cite{bbg2}. Our solutions presented in this paper can be thought as an example of this general statement. Therefore, it would  be interesting to extend our results by incorporating this general formalism in order to understand whether the most general solution including the homogeneous solutions can also be given explicitly by an all order recursive formula or not.

\begin{center}
{\bf {Acknowledgments:}}
\end{center}
We would like to thank \"{O}. F.    Day\i{} for valuable discussions. B. Yap\i{}\c{s}kan is supported by TUBITAK under BIDEP-2218.

We dedicate this work to the memory of Prof. Erdal \.In\"on\"u.



\begin{thebibliography}{99}
\addcontentsline{toc}{section}{References}




\bibitem{sw}
N.Seiberg and E.Witten, \emph{``String Theory and Noncommutative
Geometry"}, JHEP {\bf 09}, (1999)  032 [hep-th/9908142].

\bibitem{mssw}
J.~Madore, S.~Schraml, P.~Schupp and J.~Wess,
  ``Gauge theory on noncommutative spaces,''
  Eur.\ Phys.\ J.\  C {\bf 16}, 161 (2000)
  [arXiv:hep-th/0001203].


\bibitem{jssw}
B.~Jurco, S.~Schraml, P.~Schupp and J.~Wess,
  ``Enveloping algebra valued gauge transformations for non-Abelian gauge
  groups on non-commutative spaces,''
  Eur.\ Phys.\ J.\  C {\bf 17}, 521 (2000)
  [arXiv:hep-th/0006246].

\bibitem{jmssw} 
B.~Jurco, L.~Moller, S.~Schraml, P.~Schupp and J.~Wess,
  ``Construction of non-Abelian gauge theories on noncommutative spaces,''
  Eur.\ Phys.\ J.\  C {\bf 21}, 383 (2001)
  [arXiv:hep-th/0104153].

\bibitem{cjsww}
X.~Calmet, B.~Jurco, P.~Schupp, J.~Wess and M.~Wohlgenannt,
  ``The standard model on non-commutative space-time,''
  Eur.\ Phys.\ J.\  C {\bf 23}, 363 (2002)
  [arXiv:hep-ph/0111115].

\bibitem{gh}
S.~Goto and H.~Hata,
  ``Noncommutative monopole at the second order in theta,''
  Phys.\ Rev.\  D {\bf 62}, 085022 (2000)
  [arXiv:hep-th/0005101].

\bibitem{moller} 
L.~Moller,
  ``Second order of the expansions of action functionals of the  noncommutative
  standard model,''
  JHEP {\bf 0410}, 063 (2004)
  [arXiv:hep-th/0409085].

\bibitem{hakikat} 
M.~M.~Ettefaghi and M.~Haghighat,
  ``Lorentz Conserving Noncommutative Standard Model,''
  Phys.\ Rev.\  D {\bf 75}, 125002 (2007)
  [arXiv:hep-ph/0703313].

\bibitem{ana}
 A.~Alboteanu, T.~Ohl and R.~Ruckl,
  ``The Noncommutative Standard Model at $O(\theta^2)$,''
  arXiv:0707.3595 [hep-ph].

\bibitem{tw}
 J.~Trampetic and M.~Wohlgenannt,
  ``Comment on the 2nd order Seiberg-Witten maps,''
  arXiv:0710.2182 [hep-th].
  
\bibitem{ak}
T.~Asakawa and I.~Kishimoto,
  ``Comments on gauge equivalence in noncommutative geometry,''
  JHEP {\bf 9911}, 024 (1999)
  [arXiv:hep-th/9909139].

\bibitem{fidanza}
  S.~Fidanza,
  ``Towards an explicit expression of the Seiberg-Witten map at all orders,''
  JHEP {\bf 0206}, 016 (2002)
  [arXiv:hep-th/0112027].

\bibitem{wulk}
A.~Bichl, J.~Grimstrup, H.~Grosse, L.~Popp, M.~Schweda and R.~Wulkenhaar,
  ``Renormalization of the noncommutative photon self-energy to all orders  via
  Seiberg-Witten map,''
  JHEP {\bf 0106}, 013 (2001)
  [arXiv:hep-th/0104097].

\bibitem{by}
 R.~Banerjee and H.~S.~Yang,
  ``Exact Seiberg-Witten map, induced gravity and topological invariants in
  noncommutative field theories,''
  Nucl.\ Phys.\  B {\bf 708}, 434 (2005)
  [arXiv:hep-th/0404064].
  
\bibitem{su}
 E.~Ulas Saka and K.~Ulker,
  ``Dimensional reduction, Seiberg-Witten map and supersymmetry,''
  Phys.\ Rev.\  D {\bf 75}, 085009 (2007)
  [arXiv:hep-th/0701178].


\bibitem{bbg1}
  G.~Barnich, F.~Brandt and M.~Grigoriev,
  ``Seiberg-Witten maps and noncommutative Yang-Mills theories for  arbitrary
  gauge groups,''
  JHEP {\bf 0208}, 023 (2002)
  [arXiv:hep-th/0206003].

\bibitem{bbg2}
  G.~Barnich, F.~Brandt and M.~Grigoriev,
  ``Local BRST cohomology and Seiberg-Witten maps in noncommutative  Yang-Mills
  theory,''
  Nucl.\ Phys.\  B {\bf 677}, 503 (2004)
  [arXiv:hep-th/0308092].

\end{thebibliography}
\end{document}